\documentclass[11pt]{article}

\usepackage[english]{babel}

\usepackage[a4paper,top=2cm,bottom=2cm,left=3cm,right=3cm,marginparwidth=1.75cm]{geometry}
\usepackage{amsmath}
\usepackage{graphicx}
\usepackage[colorlinks=true, allcolors=blue]{hyperref}
\usepackage{url}
\usepackage{cleveref}
\crefname{figure}{Figure}{Figures}
\crefname{table}{Table}{Tables}
\crefname{section}{Section}{Sections}
\crefname{equation}{Eq.}{Eqs.}

\usepackage{graphicx}
\usepackage{graphicx}
\usepackage{empheq}

\usepackage{tikz}
\usetikzlibrary{math} %needed tikz library
\usepackage{import}
\usepackage{float}
\usepackage{placeins} % For \FloatBarrier (optional for extra float control)
\usepackage{xcolor}
\usepackage[most]{tcolorbox}
\usepackage{amsmath}
\usepackage{multirow}
\usepackage{amsmath,amsfonts,amssymb,amsthm}
\usepackage{mathtools}
\usepackage[linesnumbered,ruled]{algorithm2e}
\usepackage{mathtools}
\usepackage[super]{nth}
\usepackage[font=footnotesize,labelfont=bf]{caption}
\usepackage{subcaption}
\usepackage{graphicx}
\usepackage{wrapfig}
\usepackage{verbatim} % comments
\usepackage{longtable}
\usepackage{array}
\usepackage{supertabular}
\usepackage{longtable}
\usepackage{tabularx}
\usepackage{multirow}
\usepackage{makecell}
\usepackage{colortbl}
\usepackage{fancyhdr}
\usepackage{xcolor}
\usepackage{pgf,tikz}
\usepackage{pgfplots, pgfplotstable}

\usepackage{multido}
\usepackage{calc}
\usepackage{fp}
\usepackage{xifthen}
\usepackage{xargs}
\usepackage{etoolbox}
\DeclareMathAlphabet{\mathpzc}{OT1}{pzc}{m}{it}
\usepackage{breakcites} 
\usepackage[utf8]{inputenc}
\usepackage{helvet}
\usepackage{geometry}
\usepackage{amssymb}
\usepackage{enumitem}
\usepackage{tabularx}
\usepackage{xcolor}
\usepackage[absolute,overlay]{textpos}
\usepackage{graphicx}
\usepackage{lipsum}
\usepackage{caption}
\usepackage{multicol}
\usepackage{afterpage}
\usepackage{setspace}
\usepackage{pgffor}
\usepackage{parskip}
\usepackage{xfrac,nicefrac}
\usepackage{booktabs}
\usepackage{authblk}
\usepackage{soul}
\usepackage{mdframed}
\usepackage{leftindex}
\pgfplotsset{compat=1.18}

\usepackage[style=numeric-comp, sorting=none, giveninits=true, mincitenames=1, maxcitenames=2,maxbibnames=99]{biblatex}
\renewbibmacro{in:}{} 
\title{Reduced-order structure-property linkages for stochastic metamaterials}

\begin{document}
\author[1,2]{ Hooman Danesh\thanks {hooman.danesh@rwth-aachen.de}}
\author[2]{Maruthi Annamaraju}
\author[1]{Tim Brepols}
\author[1,3]{Stefanie Reese}
\author[2]{Surya R. Kalidindi}
\date{} % This removes the date

\affil[1]{\footnotesize{Institute of Applied Mechanics, RWTH Aachen University, Mies-van-der-Rohe-Str. 1, 52074 Aachen, Germany}}
\affil[2]{\footnotesize{George W. Woodruff School of Mechanical Engineering, Georgia Institute of Technology, 801 Ferst Drive, Atlanta, GA 30332, USA}}
\affil[3]{\footnotesize{University of Siegen, Adolf-Reichwein-Str. 2a, 57076 Siegen, Germany}}

\maketitle

\vspace{4em}

\begin{abstract}
% \setstretch{1.1}
The capabilities of additive manufacturing have facilitated the design and production of mechanical metamaterials with diverse unit cell geometries. Establishing linkages between the vast design space of unit cells and their effective mechanical properties is critical for the efficient design and performance evaluation of such metamaterials. However, physics-based simulations of metamaterial unit cells across the entire design space are computationally expensive, necessitating a materials informatics framework to efficiently capture complex structure-property relationships. In this work, principal component analysis of 2-point correlation functions is performed to extract the salient features from a large dataset of randomly generated 2D metamaterials. Physics-based simulations are performed using a fast Fourier transform (FFT)-based homogenization approach to efficiently compute the homogenized effective elastic stiffness across the extensive unit cell designs. Subsequently, Gaussian process regression is used to generate reduced-order surrogates, mapping unit cell designs to their homogenized effective elastic constant. It is demonstrated that the adopted workflow enables a high-value low-dimensional representation of the voluminous stochastic metamaterial dataset, facilitating the construction of robust structure-property maps. Finally, an uncertainty-based active learning framework is utilized to train a surrogate model with a significantly smaller number of data points compared to the original full dataset. It is shown that a dataset as small as $0.61\%$ of the entire dataset is sufficient to generate accurate and robust structure-property maps. 

\end{abstract}

\section{Introduction}
\label{sec:int}

The manufacturability of metamaterials has significantly advanced due to recent developments in additive manufacturing technologies \cite{lu2022architectural,chen2023advanced,chen2024fabrication}. While this enables a broad range of feasible geometries, it also presents a considerable challenge, as engineers must explore an increasingly complex design space to identify structures that meet specific performance requirements. Although physics-based simulations are capable of predicting the response of a given design with high accuracy, their computational cost becomes prohibitive as the design space expands. This not only limits efficient design space exploration, but also renders inverse design approaches largely impractical. Consequently, there is a critical need for accurate and efficient surrogate models that can establish reliable mappings between the geometry of cellular structures and their effective properties (i.e., structure-property linkages).

Data-driven machine learning approaches have been widely employed to establish structure-property linkages for various metamaterial families and diverse mechanical properties \cite{tajalsir2022numerical,luan2023data,ben2023gam,danesh2024fft,mishra2024latticeml,wang2020deep,meyer2022graph,bastek2023inverse,xiao2024machine,liu2025machine}. These approaches range from different ensemble learning algorithms \cite{tajalsir2022numerical,luan2023data,ben2023gam,danesh2024fft,mishra2024latticeml} to various neural network architectures \cite{wang2020deep,meyer2022graph,bastek2023inverse,xiao2024machine,liu2025machine}. Focusing on different approaches is beyond the scope of this paper, and interested readers are referred to recently published review articles on the application of machine learning for metamaterials \cite{zheng2023deep,cerniauskas2024machine,song2024artificial,lee2024data}. Such data-driven approaches typically require large datasets to train accurate and robust models, as they rely solely on empirical patterns rather than physical knowledge. Although physics-informed approaches \cite{raissi2019physics,karniadakis2021physics} may reduce the data required, they can introduce challenges, such as the complexity of formulating the underlying physics, training instabilities, or computational overhead \cite{krishnapriyan2021characterizing,wang2022and,cuomo2022scientific,farea2024understanding}. Gaussian process regression (GPR) \cite{seeger2004gaussian,williams2006gaussian,schulz2018tutorial,wang2023intuitive}, as a nonparametric approach that provides uncertainty estimates alongside predictions, offers an alternative to establish robust models with significantly less data. Although the application of GPR for building structure-property relationships in metamaterials remains underexplored, this approach has been demonstrated in a few studies to train models effectively using small datasets \cite{wang2018analysis,wang2018optimization,song2023free,challapalli2023inverse}. 

When it comes to establishing structure–property maps for metamaterial unit cells, a critical consideration is how the design geometry is represented as the input of the model. Although many classes of metamaterials can be effectively described using geometric parameters (e.g., strut thickness, void diameter, volume fraction, rotational angle, etc.) \cite{lee2024data}, such parametrizations inherently constrain the design space to a limited set of idealized configurations. In practice, however, the space of manufacturable unit cells is considerably broader, encompassing stochastic and randomly generated architectures \cite{bostanabad2019globally,wang2020data,xue2020machine,bastek2023inverse}, which are not easily captured by geometric descriptors. Furthermore, fabrication processes inevitably introduce deviations from the intended design, including dimensional inaccuracies and local defects, all of which can significantly affect material behavior \cite{morris2018design,du2020effects,xiao2020multi,dastani2023effect,cuan2023fused}. Therefore, surrogate models that rely exclusively on idealized geometric parameters may overlook critical factors influencing material performance. This thereby highlights the need for surrogate models that incorporate the spatial geometry of the unit cell, since parametric representations may fail to capture the complexity of random architectures and overlook critical fabrication imperfections.

While metamaterial unit cells can be characterized using various microscopy and tomography techniques \cite{morris2018design,germain20183d,cuan2023fused}, it should be noted that these techniques only provide a single instantiation of the material structure and not the material structure itself. This is because multiple realizations of the material structure can be obtained from a single physical sample, each exhibiting inherent variability in the extracted statistical features \cite{ kalidindi2011microstructure,kalidindi2015hierarchical,kalidindi2020feature}. Such variability introduces uncertainty arising from both the characterization process and sample preparation. In addition to these uncertainties, there is also an arbitrary aspect in defining a metamaterial unit cell due to the absence of a natural origin. Any region that repeats periodically to reconstruct the full structure can serve as a valid unit cell, with no single fixed choice \cite{li2019representative}. Nevertheless, these different unit cell representations describe the same physical material and should, in principle, yield consistent predictions in structure–property linkages. Together, these sources of variability highlight the need for a stochastic framework for microstructure quantification that considers the material structure as a random process.

The recently developed materials knowledge systems (MKS) framework \cite{kalidindi2011microstructure,kalidindi2015hierarchical,kalidindi2020feature,brough2017materials,yabansu2019application,marshall2021autonomous,yabansu2020digital,hashemi2022feature,montes2018reduced,yabansu2017extraction,latypov2017data,latypov2019materials,al2012multi} quantifies material microstructure as a random process using 2-point spatial correlations \cite{fullwood2008strong,niezgoda2008delineation,fullwood2010microstructure,niezgoda2011understanding,niezgoda2013novel}. This higher-order statistical representation captures spatial information in a simple yet effective manner and bypasses the need for manual feature selection in establishing structure–property linkages. However, these 2-point statistics yield a large set of features that are unwieldy for establishing structure-property maps. Therefore, MKS leverages the principal component analysis (PCA) \cite{suh2002application,bro2014principal,jolliffe2016principal} to achieve a reduced set of features for the low-dimensional representation of spatial correlations. Reduced-order structure-property linkages can then be established using a wide range of available regression techniques. However, it has been demonstrated in previous works \cite{yabansu2019application,yabansu2020digital,marshall2021autonomous,hashemi2022feature} that GPR provides key benefits by employing a nonparametric modeling approach and rigorous treatment of prediction uncertainty. 

In most practical applications, generating data for structure–property modeling, such as through high-fidelity simulations or physical experiments, is expensive and resource intensive, necessitating frameworks that minimize data needs while maintaining model accuracy. Active learning addresses this challenge by iteratively selecting the most informative samples to build effective surrogate models with limited data \cite{settles2009active}. Various approaches exist to identify these samples (i.e., sampling strategies) \cite{settles2009active}, among which uncertainty sampling has proven to be a simple yet effective strategy for GPR models, owing to the inherent uncertainty quantification provided by GPR \cite{liu2024active,buzzy2025active,ozbayram2025batch}. By prioritizing samples where the model uncertainty is highest, uncertainty sampling ensures that each new training point maximally improves the model, enhancing the fidelity of the surrogate with as little data as possible \cite{lewis1995sequential,settles2009active}.

Although the MKS framework has already been applied to composites with different ranges of stiffness contrast \cite{al2012multi,latypov2017data,latypov2019materials,fernandez2019comparative}, to the best of our knowledge, it has not yet been used to model the mechanical response of materials with infinite contrast, such as the mechanical metamaterials in the present work. To this end, we employ the MKS framework to establish reduced-order structure–property linkages for an existing stochastic metamaterial dataset \cite{bastek2023inverse,bastek2023data}. The primary quantity of interest in this study is the $C_{11}$ component ($C_{1111}$ in full tensorial notation) of the effective homogenized elastic stiffness, chosen as a representative example, while the approach is generalizable to other components as well. In order to compute this homogenized elastic constant, we perform physics-based simulations using an efficient fast Fourier transform (FFT)-based homogenization method \cite{lucarini2022adaptation,danesh2024fft}. Motivated by previous works \cite{yabansu2020digital,hashemi2022feature,montes2018reduced} showcasing higher achievable fidelity by incorporating the statistics of the interface between the phases present in the microstructure, we also extract the solid–void interface for each metamaterial unit cell to investigate its influence on the performance of the surrogate model. The reduced-order representation of the metamaterial unit cells is then obtained by computing the 2-point spatial correlations for both the solid and the interface phases, followed by the application of PCA on different combinations of the computed 2-point statistics. GPR is subsequently used to establish the structure–property maps by linking the reduced set of features obtained from different combinations of the 2-point statistics to the homogenized elastic constant. Our findings show that, although modest, the inclusion of the interface information improves the accuracy of the surrogate when combined with solid–solid statistics. Once the most accurate model is selected, we adopt an uncertainty-based active learning approach to identify the smallest subset of data necessary for generating structure–property maps with optimal predictive performance. We demonstrate that a surrogate model trained on approximately $0.61\%$ of the dataset can achieve nearly the same accuracy as a model trained on the full dataset.

\section{Methodology}

\subsection{Stochastic metamaterial dataset}
\label{sec:dataset}

\begin{figure}[H]
    \centering
    \includegraphics[width=0.5\textwidth,keepaspectratio]{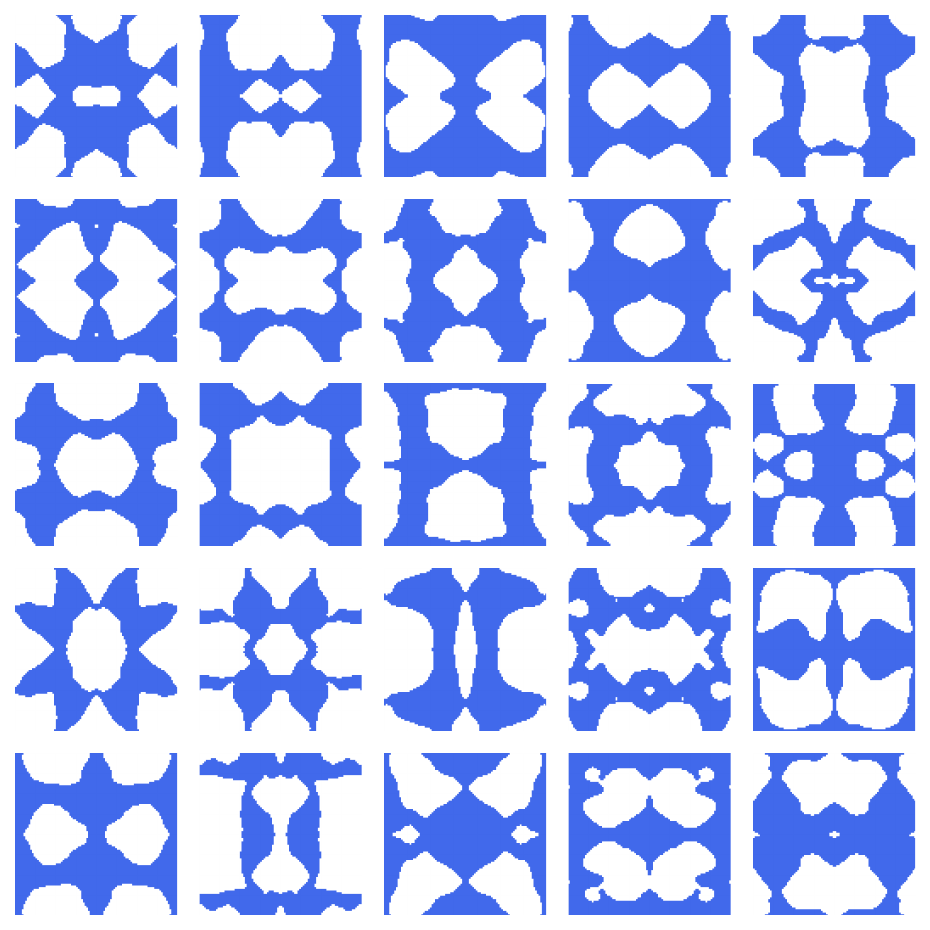}
    \caption{Example unit cells from the dataset of periodic 2D stochastic metamaterials.}
    \label{fig:example_ucs}
\end{figure}

To leverage the capabilities of the MKS framework, a voluminous dataset of stochastic metamaterial unit cells is beneficial. Here, the dataset of periodic 2D stochastic cellular structures generated by \textcite{bastek2023inverse,bastek2023data} is employed, consisting of 53,019 microstructures represented on a $96\times96$ pixel grid. Each structure is created by sampling a Gaussian random field, which is then binarized using a randomly selected threshold to distinguish solid from void regions. To ensure structural connectivity and periodicity, only samples with sufficient material presence along all boundaries are retained. The accepted structures are then sequentially mirrored along both horizontal and vertical directions to produce periodic unit cells. As a pixel-based representation, this dataset is particularly well suited for the MKS framework. A randomly selected subset of $25$ unit cells from this large dataset is shown in \cref{fig:example_ucs}.

This generation approach, which imposes minimal design constraints apart from periodicity, results in a diverse collection of microstructures. The resulting structures span a broad range of relative densities, from $0.30$ to $0.68$, further contributing to the dataset geometric diversity. Nevertheless, to quantitatively assess the diversity within the dataset, the average pairwise Euclidean distance between all structures is computed. The resulting average is $66$. Given that the maximum possible distance for binary $96\times96$ images is $96$, this yields a normalized diversity score of $0.69$ (with $1$ being the theoretical maximum). This indicates a relatively high level of dissimilarity between structures, with unit cells differing by approximately $69\%$ on average. To further investigate the diversity of the dataset, k-means clustering with a large number of $500$ clusters is applied. It is found that, despite using such a high number of clusters, the coefficient of variation in the cluster sizes is $0.19$. This suggests that the dataset is not concentrated in just a few clusters but is instead broadly distributed across the design space, further highlighting the diversity of the considered dataset.

\subsection{Fast Fourier transform (FFT)-based homogenization}
\label{sec:FFT}

The FFT-based homogenization scheme was first introduced by \textcite{moulinec1994fast, moulinec1998numerical} to study the homogenized effective response of both linear and nonlinear composites. Reformulating the homogenization problem as the convolution of local strain fluctuations with the Green's function of a reference elastic medium leads to the periodic Lippmann–Schwinger equation. Owing to the availability of the Green's function in closed form in Fourier space, and the advantage of transforming the convolution into algebraic products in the Fourier domain, the Lippmann–Schwinger equation can be efficiently solved using FFT in combination with fixed-point iterations (see, e.g., \cite{schneider2021review,lucarini2021fft} for review).

However, the convergence of the basic scheme described above depends on the stiffness contrast between the phases present in the microstructure, making it unsuitable for materials with very high or infinite contrast, such as the metamaterial unit cells investigated in the present work \cite{danesh2023challenges}. Among the various approaches developed to address this limitation \cite{michel2001computational,brisard2010fft,monchiet2012polarization,schneider2016computational}, \textcite{lucarini2022adaptation} effectively extended FFT-based homogenization to metamaterials by employing the Galerkin FFT method \cite{vondvrejc2014fft,vondvrejc2015guaranteed}, combined with a rotated finite difference grid \cite{willot2015fourier} and the MINRES solver. This framework was also adopted in our previous work \cite{danesh2024fft} to generate high-fidelity datasets by computing the homogenized elastic constants of auxetic unit cells.

The present work employs the same methodology \cite{danesh2024fft} to compute the components of the effective stiffness tensor $\mathbb{C}$ for the large dataset of metamaterial unit cells introduced in \cref{sec:dataset}. As the full formulation and implementation details are already provided in \cite{danesh2024fft}, they are not repeated here. Instead, we briefly recall the core idea of the perturbation-based approach \cite{idrissi2022multiparametric,gierden2022review} used to extract the effective stiffness components. The homogenized elastic stiffness tensor $\mathbb{C}$ is defined as

\begin{equation}
\mathbb{C} = \frac{\partial\bar{\boldsymbol{\sigma}}}{\partial\bar{\boldsymbol{\varepsilon}}},
\end{equation}

where $\bar{\boldsymbol{\varepsilon}}$ and $\bar{\boldsymbol{\sigma}}$ denote the macroscopic average strain and stress tensors, respectively. The components of the homogenized effective elastic stiffness $C_{ijkl}$ are computed by applying an independent macroscopic strain perturbation $\bar{\varepsilon}_{kl} = \beta$, with all other components set to zero. The resulting average stress response $\bar{\sigma}_{ij(kl)}$ is then used to evaluate

\begin{equation}
C_{ijkl} = \frac{\bar{\sigma}_{ij(kl)}}{\beta},
\end{equation}

where $\beta$ is a constant strain perturbation. Considering linear elasticity, the specific value of $\beta$ is irrelevant, and the unperturbed state is taken as zero strain. While three (in 2D) or six (in 3D) independent perturbations are required to compute all independent components of the effective stiffness tensor, we focus in this work solely on the $C_{1111}$ component (denoted $C_{11}$ in reduced notation), which can be obtained from a single perturbation $\bar{\varepsilon}_{11} = \beta$ as

\begin{equation}
C_{11} = C_{1111} = \frac{\bar{\sigma}_{11(11)}}{\beta}.
\end{equation}

\subsection{2-point spatial correlations}
\label{sec:2ps}

As discussed in \cref{sec:int} and in previous works \cite{kalidindi2011microstructure,kalidindi2015hierarchical,kalidindi2020feature}, obtaining a low-dimensional representation of the material microstructure (i.e., metamaterial unit cells in our case) requires the computation of 2-point spatial correlation functions prior to the application of PCA. To achieve this, the discretized microstructure array $m_{\mathbf{s}}^{h}$ is defined as the volume fraction of the local state $h$ at the discretized spatial bin  $\mathbf{s} \in \mathcal{S}$ (i.e., voxel in 3D or pixel in 2D). The spatial index $\mathbf{s}$ can be represented as $\mathbf{s} = (s_1, s_2)$ in 2D and $\mathbf{s} = (s_1, s_2, s_3)$ in 3D, corresponding to the position in the discretized microstructure. Considering that each spatial bin contains only a single distinct local state, the microstructure array $m_{\mathbf{s}}^{h}$ can take only the binary values 0 or 1. Subsequently, the discretized 2-point correlation functions, denoted as $f^{hh'}_{\mathbf{r}}$, quantify the probability of simultaneously observing the local states $h$ and $h'$ at two spatial locations separated by a vector $\mathbf{r} = (r_1, r_2, r_3)$, and are mathematically formulated as \cite{kalidindi2015hierarchical,kalidindi2020feature,fullwood2008strong,niezgoda2008delineation,fullwood2010microstructure,niezgoda2011understanding,niezgoda2013novel}

\begin{equation}
\label{eq:2ps}
    f^{hh'}_{\mathbf{r}} = \dfrac{1}{S_\mathbf{r}} \sum_{\mathbf{s} \in \mathcal{S}} m_{\mathbf{s}}^{h}\, m_{\mathbf{s}+\mathbf{r}}^{h'},
\end{equation}

where $S_\mathbf{r}$ is the total number of trials accessible for the valid placement of the vector $\mathbf{r}$. In this setting, $ f^{hh'}_{\mathbf{r}}$ is referred to as auto-correlation when $h=h'$ and as cross-correlation when $h \neq h'$. It is worth noting that for periodic microstructures, such as the cellular structures in the present work, $S_\mathbf{r}$ is equal to the total number of spatial bins $|\mathcal{S}|$ (i.e., $S_\mathbf{r} = |\mathcal{S}|$) \cite{kalidindi2015hierarchical,cecen2016versatile}. Eventually, taking advantage of the convolution theorem, the 2-point statistics for periodic microstructures can readily be computed using FFT as \cite{fullwood2008microstructure,fullwood2008strong,cecen2016versatile}

\begin{equation}
\label{eq:2psfft}
    f^{hh'}_{\mathbf{r}} = \dfrac{1}{|\mathcal{S}|} \mathcal{F}^{-1}\left[ \left(\mathcal{F}\left[{m}^{h}_{\mathbf{s}} \right]\right)^{*} \, \mathcal{F}\left[ m^{h'}_{\mathbf{s}} \right] \right],
\end{equation}

where $(\cdot)^*$ denotes the complex conjugate, and $\mathcal{F}[\cdot]$ and $\mathcal{F}^{-1}[\cdot]$ represent the forward and inverse Fourier transforms, respectively. It is also worth noting that, even in the absence of periodicity, FFT algorithms can still be applied to compute 2-point correlations by leveraging a padding strategy \cite{kalidindi2015hierarchical,cecen2016versatile}.

\subsection{Principal component analysis (PCA)}
\label{sec:PCA}

The computation of the 2-point correlations through \cref{eq:2ps,eq:2psfft} results in an extensive and unmanageable set of features. It is therefore practical to apply dimensionality reduction techniques to obtain a reduced yet informative feature space. PCA \cite{suh2002application,bro2014principal,jolliffe2016principal} has been shown to be effective in capturing a significantly smaller set of salient features ideal for establishing structure–property linkages \cite{yabansu2020digital,marshall2021autonomous,hashemi2022feature,yabansu2017extraction,montes2018reduced,latypov2017data,latypov2019materials}. PCA is a linear distance-preserving transformation that identifies orthogonal linear combinations of the original features, which are ordered by the amount of variance they explain in the dataset, from highest to lowest. 

Let $j \in \{1,2, \dots, J\}$ index the members of the ensemble of microstructures, and let $f^{(j)}_{r}$, with $r \in \{1,2, \dots, R\}$, represent the high-dimensional vectorized representation of all 2-point statistics deemed important to generate structure–property maps for the $j$-th microstructure, where $J$ and $R$ denote the total number of microstructures and features, respectively. The principal component (PC) representation of this microstructure exemplar can then be written as \cite{kalidindi2015hierarchical,kalidindi2011microstructure,niezgoda2011understanding,niezgoda2013novel}

\begin{equation}
\label{eq:PCA}
    {f}^{(j)}_{r} = \sum_{i=1}^{\min(J-1,R)} \alpha^{(j)}_{i} \varphi_{ir} + \bar{f}_{r},
\end{equation}

where $\alpha^{(j)}_{i}$ denotes the $i$-th PC score of the $j$-th microstructure, $\varphi_{ir}$ are the PC basis vectors, and $\bar{f}_{r}$ is the ensemble average. As shown in \cref{eq:PCA}, PCA identifies up to $\min(J - 1, R)$ orthogonal basis vectors (i.e., PCs) in the feature space. However, due to the descending order of variance captured by the PC scores, it is possible to objectively truncate the series in \cref{eq:PCA} to obtain a low-dimensional representation without significant loss of information. It has been demonstrated in previous studies \cite{yabansu2020digital,marshall2021autonomous,hashemi2022feature,yabansu2017extraction,montes2018reduced,latypov2017data,latypov2019materials} that only a limited number of PC scores are sufficient to achieve a large portion of the explained variance from the entire dataset, resulting in high-fidelity structure-property linkages. Assuming $\hat{R}$ denotes the number of retained components, it typically holds that $\hat{R} \ll \min (J-1,R)$.

\subsection{Gaussian process regression (GPR)}
\label{sec:gpr}

After extracting the salient features of the microstructures and obtaining the low-dimensional representation, the mapping of input features to the desired output can be performed through various model building approaches. Among others, GPR has proven to be a powerful tool to establish structure–property linkages in previous works \cite{yabansu2019application,yabansu2020digital,marshall2021autonomous,hashemi2022feature}. This nonparametric approach not only predicts the outputs of interest but also provides the uncertainty associated with the predictions \cite{seeger2004gaussian,williams2006gaussian,schulz2018tutorial,wang2023intuitive}, which offers additional insight and becomes especially useful when employing active learning approaches (see \cref{sec:AL}).

Assuming a zero-mean Gaussian process (GP), the output function $y$ is modeled by a multivariate normal distribution $\mathcal{N}$ as \cite{williams2006gaussian}

\begin{equation}
    y(\mathbf{x}) \sim \mathcal{N} \left( 0, k(\mathbf{x}, \mathbf{x}') \right),
\end{equation}

where $\mathbf{x}$ denotes the vector of input features with the dimension $D$, and $k(\mathbf{x}, \mathbf{x}')$ is the covariance function of the GP. The automatic relevance determination squared exponential (ARD-SE) kernel has proven to be a suitable choice for the GP covariance due to its flexibility in assigning a separate hyperparameter to each input feature. The mathematical expression for the ARD-SE kernel is given by \cite{williams2006gaussian}:

\begin{equation}
    k(\mathbf{x},\mathbf{x}') = \sigma_f^2 \exp \left( -\frac{1}{2} \sum_{d=1}^{D} \frac{(x_d - x_d')^2}{\ell_d^2} \right) + \sigma_n^2 \delta_{\mathbf{xx}'}.
\end{equation}

Here, the scaling factor of the output variance $\sigma_f$, the noise factor $\sigma_n$, and the characteristic length-scale $\ell_d$ corresponding to the input variable $x_d$ comprise the set of hyperparameters for the kernel function $k(\mathbf{x},\mathbf{x}')$, while $\delta_{\mathbf{xx}'}$ represents the Kronecker delta function. Let $\mathbf{y}$ and $\mathbf{y}_*$ denote the observed training and unseen test output vectors with $N$ and $N_*$ data points, respectively. The joint distribution of the training outputs $\mathbf{y}$ and the test outputs $\mathbf{y}_*$ is then expressed as \cite{williams2006gaussian}

\begin{equation}
\label{eq:jointdist}
    \begin{bmatrix}
        \mathbf{y} \\ \mathbf{y}_*
        \end{bmatrix}
        \sim \mathcal{N} \left( \mathbf{0},
        \begin{bmatrix}
        \mathbf{K} & \mathbf{K}_* \\
        \mathbf{K}_*^\top & \mathbf{K}_{**}
    \end{bmatrix} \right).
\end{equation}

In \cref{eq:jointdist}, $\mathbf{K}=k(\mathbf{X,X})$, $\mathbf{K}_*=k(\mathbf{X},\mathbf{X}_*)$, and $\mathbf{K}_{**}=k(\mathbf{X}_*,\mathbf{X}_*)$, where $\mathbf{X}$ and $\mathbf{X}_*$ denote the $N \times D$ and $N_* \times D$ input matrices for the training and test data, respectively. Leveraging the conditional Gaussian distribution, the mean $\boldsymbol{\mu}_*$ and variance $\boldsymbol{\Sigma}_*$ (i.e., uncertainty) of the predictive distribution at the test points are obtained by \cite{williams2006gaussian}

\begin{align}
    \boldsymbol{\mu}_* &= \mathbf{K}_*^{\top} \mathbf{K}^{-1} \mathbf{y}, \\
    \boldsymbol{\Sigma}_* &= \mathbf{K}_{**} - \mathbf{K}_*^{\top} \mathbf{K}^{-1} \mathbf{K}_*.
\end{align}

The predictive accuracy of the GPR model is highly sensitive to the choice of hyperparameters. Let $\boldsymbol{\theta}=\left( \sigma_f, \sigma_n, \ell_1, \ell_2, \dots, \ell_D \right)$ denote the vector of all hyperparameters characterizing the GP model. The optimal hyperparameters $\hat{\boldsymbol{\theta}}$ are determined from the training data by minimizing the negative log likelihood function $\log p(\mathbf{y} | \mathbf{X}, \boldsymbol{\theta})$ as

\begin{equation}
\label{eq:opthyp}
    \hat{\boldsymbol{\theta}} = \arg\min_{\boldsymbol{\theta}} \left(-\log p(\mathbf{y} \mid \mathbf{X}, \boldsymbol{\theta}) \right),
\end{equation}

where the log marginal likelihood $\log p(\mathbf{y} | \mathbf{X}, \boldsymbol{\theta})$ is given by \cite{williams2006gaussian}

\begin{equation}
\label{eq:mll}
    \log p(\mathbf{y} | \mathbf{X}, \boldsymbol{\theta}) = -\frac{1}{2} \mathbf{y}^\top \mathbf{K}^{-1} \mathbf{y} - \frac{1}{2} \log \det(\mathbf{K}) - \frac{N}{2} \log 2\pi.
\end{equation}

Finally, it is worth noting that, considering that our main output in the present study is the homogenized stiffness component $C_{11}$, we employ the mean absolute stiffness error (MAE) on the unseen test data as the main error measure to evaluate the predictive performance of the model. The MAE is defined as

\begin{equation}
\label{eq:MAE}
    \text{MAE} = \frac{1}{N_*} \sum_{n=1}^{N_*} \left| C^{(n)}_{11} - \hat{C}^{(n)}_{11} \right|,
\end{equation}

where $n \in \{1,2, \dots, N_*\}$ enumerates the samples (i.e., metamaterial unit cells) in the test dataset, $C_{11}$ is the actual value obtained from physics-based simulations, and $\hat{C}_{11}$ denotes the predicted value by the GPR model.

\subsection{Active learning}
\label{sec:AL}

To establish structure–property linkages, it is essential to label the data, i.e., to compute the property of interest for each metamaterial unit cell and assign it accordingly. However, when the amount of unlabeled data is large or the labeling process is expensive (e.g., due to high computational cost of simulations or time-consuming and costly experiments), it becomes crucial to develop models that can achieve high performance with as little labeled data as possible.

Active learning is an effective approach in such scenarios, where the algorithm interactively queries new data points to be labeled. Several strategies exist for generating candidate data points for labeling, including membership query synthesis, stream-based sampling, pool-based sampling, etc. \cite{settles2009active}. The present study adopts the pool-based sampling strategy, as the dataset is readily available. In pool-based sampling, candidate data points are known \textit{a priori}, unlike the other methods where candidates are generated dynamically during the learning process. It is therefore essential that the candidate pool sufficiently covers the entire input space, ensuring that all possible input microstructures are taken into account.

Another key aspect of active learning is the query strategy, which determines how new training data points are selected for labeling. A wide range of strategies has been proposed in the literature, such as random sampling, query-by-committee, and uncertainty sampling, among others \cite{settles2009active}. Given that GPR provides an estimate of prediction uncertainty, uncertainty sampling is particularly well suited for this work. In this study, a simple yet effective approach is used: at each active learning iteration, the data point with the highest predictive uncertainty (i.e., maximum variance) is selected from the candidate pool and added to the training set. This criterion has been shown to be effective \cite{liu2024active,buzzy2025active,ozbayram2025batch}, especially when the model noise is relatively uniform across the input space \cite{settles2009active}.

Finally, it is important to define a clear stopping criterion for the active learning process. Without such a criterion, the algorithm may continue querying new data points unnecessarily, leading to inefficiencies in computational or experimental resources. Typically, active learning iterations continue until a predefined labeling budget is exhausted or a desired level of model accuracy is reached. In practice, stopping criteria can also include convergence of model performance metrics (e.g., error measures) or stabilization of prediction uncertainty, indicating that adding more data no longer yields meaningful improvements. As mentioned in \cref{sec:gpr}, the MAE is employed as the error metric to evaluate the predictive performance of the model. Therefore, we adopt the following stopping criterion, defined by the relative change in MAE over a sliding window of recent active learning iterations:

\begin{equation}
\label{eq:stop}
    \frac{1}{Q} \sum_{q=I-Q+1}^{I} \left| \frac{\text{MAE}(\hat{\boldsymbol{\theta}}_{q}) - \text{MAE}(\hat{\boldsymbol{\theta}}_{q-1})}{\text{MAE}(\hat{\boldsymbol{\theta}}_{q-1})} \right| < \epsilon.
\end{equation}

In \cref{eq:stop}, $\hat{\boldsymbol{\theta}}_q$ represents the optimized model hyperparameters at the $q$-th active learning iteration, while $I$ denotes the total number of completed iterations. The parameter $Q$ is a user-defined value specifying the number of recent iterations over which the relative change in MAE is averaged. The stopping criterion is met when the average relative change in MAE over the last $Q$ iterations falls below a predefined threshold $\epsilon$, indicating convergence in model performance and diminishing benefit from further sampling.

\newpage
\section{Results and discussion}

\subsection{Feature engineering}

Following the MKS framework, we employ the workflow of 2-point correlation functions and PCA, as formulated in \cref{sec:2ps,sec:PCA}, to obtain a reduced set of salient features for the metamaterial dataset described in \cref{sec:dataset}. To do so, we start by labeling the local states $h$ present in the microstructure. Since we are dealing with metamaterial unit cells, we initially consider two local states $h \in \{0,1\}$, where the void and solid phases are denoted by $h=0$ and $h=1$, respectively. In this setting, $m_{\mathbf{s}}^{0}$ represents the volume fraction of the void phase at pixel $\mathbf{s}$, while $m_{\mathbf{s}}^{1}$ refers to the volume fraction of the solid phase. As the metamaterial unit cells at hand are eigen microstructures (i.e., only a single local state is present at each spatial bin), the discretized microstructure array $m_{\mathbf{s}}^{h}$ can only take binary values, 0 or 1.

\begin{figure}[H]
    \centering
    \includegraphics[width=0.9\textwidth,keepaspectratio]{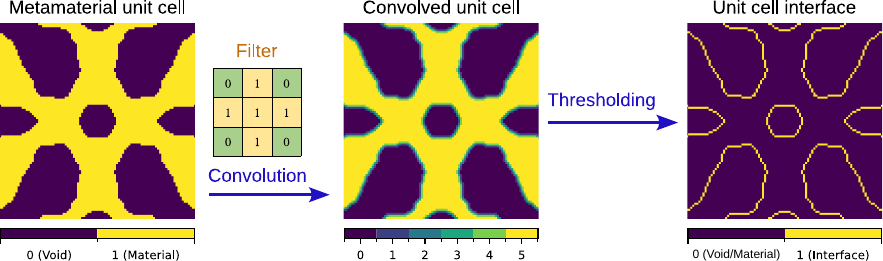}
    \caption{Interface extraction workflow: the solid microstructure array $m_{\mathbf{s}}^{1}$ (left) is convolved with a 5-point kernel to produce the filtered field $m_{\mathbf{s}}^{1,\text{conv}}$ (middle), where each pixel value indicates the number of neighboring pixels in the solid phase. The interface microstructure array $m_{\mathbf{s}}^{2}$ (right) is obtained by labeling void pixels that are adjacent to at least one solid pixel.}
    \label{fig:interface}
\end{figure}

Previous studies have shown the effect of the interface between different phases in the microstructure on the accuracy of surrogate models for various physical phenomena, such as permeability \cite{yabansu2020digital}, charge transport \cite{hashemi2022feature}, and damage initiation \cite{montes2018reduced}. Motivated by these works, we use a 2D convolution operation to extract the interface pixels between the solid and void phases in metamaterial unit cells and use this as a third material state descriptor at the pixel scale. An illustrative example of the interface extraction is shown in \cref{fig:interface}. The left plot shows the pixelized microstructure, where pixels with value 1 represent the solid phase and those with value 0 correspond to the void phase. This binary field is convolved with a 5-point kernel $\phi_{\mathbf{c}}$ as

\begin{equation}
    m_{\mathbf{s}}^{1,\text{conv}} = \sum_{\mathbf{c} \in \mathcal{C}} \phi_{\mathbf{c}} \, m_{\mathbf{s} + \mathbf{c}}^1,
    \quad
    \phi_{\mathbf{c}} = 
        \begin{bmatrix}
        0 & 1 & 0 \\
        1 & 1 & 1 \\
        0 & 1 & 0
        \end{bmatrix},
\end{equation}

where $m_{\mathbf{s}}^{1,\text{conv}}$ is the convolved microstructure array, $\mathbf{c}$ denotes relative coordinates in the kernel, and $\mathcal{C} = \{(0,0), (0,1), (0,-1), (1,0), (-1,0)\}$. 
%includes the non-zero entries of $\phi_{\mathbf{c}}$. 
The convolution is performed assuming periodic boundary conditions on the microstructure, consistent with the assumptions implicit in the FFT-based homogenization method described earlier. The resulting convolved field, shown in the middle plot of \cref{fig:interface}, contains integer values between 0 and 5, each representing the number of solid phase pixels within a 5-point stencil, including the central pixel and its four immediate neighbors. We then define the interface using a simple criterion, based on which a single pixel on the void side adjacent to at least one solid pixel is labeled as an interface pixel. Formally, this is implemented through a thresholding operation:

\begin{equation}
\label{eq:thresh}
    m_{\mathbf{s}}^{2} = \chi_{\left\{ m_{\mathbf{s}}^1 = 0 \right\} \cap \left\{ m_{\mathbf{s}}^{1,\text{conv}} > 0 \right\}}(\mathbf{s}).
\end{equation}

Here, $m_{\mathbf{s}}^{2}$ denotes the interface microstructure array, where $m_{\mathbf{s}}^{2} = 1$ identifies the interface pixels, while $m_{\mathbf{s}}^{2} = 0$ elsewhere (i.e., void or solid phases). The symbol $\chi_{\mathcal{A}}(\mathbf{s})$ denotes the indicator function of a set $\mathcal{A}$, which returns 1 if $\mathbf{s} \in \mathcal{A}$ and 0 otherwise. Additionally, the operator $\cap$ denotes the intersection of two sets, meaning that $\mathbf{s}$ must satisfy both conditions simultaneously. The right plot in \cref{fig:interface} shows the extracted interface for the example unit cell. This operation introduces a third local state, $h = 2$, representing the interface, thereby extending the set of local states to $h \in \{0,1,2\}$.

\begin{figure}[H]
    \centering
    \includegraphics[width=0.9\textwidth,keepaspectratio]{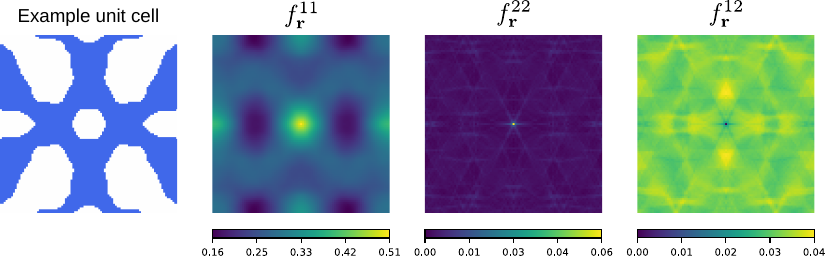}
    \caption{2-point spatial correlations for an example unit cell, including the solid–solid auto-correlation $f^{11}_{\mathbf{r}}$ (left), the interface–interface auto-correlation $f^{22}_{\mathbf{r}}$ (middle), and the solid–interface cross-correlation $f^{12}_{\mathbf{r}}$ (right).}
    \label{fig:2ps}
\end{figure}

After defining the local states of the metamaterial unit cells, we compute the most informative 2-point spatial correlations. For a microstructure with $H$ local states, prior work \cite{niezgoda2008delineation} has shown that only $H-1$ of the $H^2$ possible 2-point correlations are independent. However, the optimal combination for establishing high-fidelity linkages is not known \textit{a priori}, as potential nonlinear dependencies among these correlations may not be captured by PCA \cite{yabansu2020digital}. For the present metamaterial unit cells with three local states (i.e., void, solid, and interface), only two of the nine possible sets of correlations (three auto-correlations and six cross-correlations) are independent. However, in a previous study \cite{yabansu2017extraction}, it was shown that considering a combination of three sets of 2-point correlations is beneficial when dealing with microstructures with three local states, where one of these sets must be a cross-correlation. Building on these observations, we select three sets of correlations for our analysis: the solid–solid auto-correlation $f^{11}_{\mathbf{r}}$, the interface–interface auto-correlation $f^{22}_{\mathbf{r}}$, and the solid–interface cross-correlation $f^{12}_{\mathbf{r}}$. The computed 2-point correlations for an example metamaterial unit cell are shown in \cref{fig:2ps}. A notable feature of the auto-correlation maps is that their value at the zero shift vector $\mathbf{r} = \mathbf{0}$ reflects the volume fraction of the associated phase. More specifically, in \cref{fig:2ps}, the values of the central pixels in the solid and interface auto-correlation maps (i.e., $f^{11}_{\mathbf{0}}$ and $f^{22}_{\mathbf{0}}$) correspond to the volume fractions of the solid and interface phases, respectively.

As mentioned in earlier works \cite{yabansu2017extraction, yabansu2020digital, hashemi2022feature, montes2018reduced}, it is necessary to rescale the computed 2-point statistics prior to dimensionality reduction, as PCA gives greater weight to features with higher numerical values due to their stronger influence on the overall variance. Therefore, inspired by \cite{hashemi2022feature}, we rescale the 2-point correlations in a way that each set exhibits the same variance. To achieve this, we compute the mean and, consequently, the standard deviation of each set of correlations across all spatial bins and unit cells in the ensemble as follows:

\begin{equation}
\begin{aligned}
    \mu^{hh'} &= \frac{1}{J} \frac{1}{|\mathcal{S}|} \sum_{j=1}^{J} \sum_{\mathbf{r}} {}^{(j)}f^{hh'}_{\mathbf{r}}, \\[1em]
    \sigma^{hh'} &= \sqrt{\frac{1}{J} \frac{1}{|\mathcal{S}|} \sum_{j=1}^{J}  \sum_{\mathbf{r}} \left( {}^{(j)}f^{hh'}_{\mathbf{r}} - \mu^{hh'} \right)^2 },
\end{aligned}
\end{equation}

where $\mu^{hh'}$ and $\sigma^{hh'}$ denote the mean and standard deviation across all bins and unit cells corresponding to the 2-point correlation set $f^{hh'}_{\mathbf{r}}$. The rescaled 2-point correlation of the $j$-th unit cell in the ensemble ${}^{(j)}\tilde{f}^{hh'}_{\mathbf{r}}$ is then computed as

\begin{equation}
\label{eq:scale2ps}
    {}^{(j)}\tilde{f}^{hh'}_{\mathbf{r}} = \left( \frac{\sigma^{11}}{\sigma^{hh'}} \right) {}^{(j)}f^{hh'}_{\mathbf{r}}.
\end{equation}

In \cref{eq:scale2ps}, $\sigma^{11}$ denotes the standard deviation corresponding to the solid auto-correlation $f^{11}_{\mathbf{r}}$. In this way, the rescaled correlations $\tilde{f}^{22}_{\mathbf{r}}$ and $\tilde{f}^{12}_{\mathbf{r}}$ are adjusted to have the same standard deviation as that of the solid auto-correlation $f^{11}_{\mathbf{r}}$. Although $f^{11}_{\mathbf{r}}$ remains unchanged during this rescaling, we still use the notation $\tilde{f}^{11}_{\mathbf{r}}$ for consistency. The three computed 2-point spatial correlations result in seven possible combinations. However, for the reasons mentioned above, we consider the following three combinations: $\tilde{f}^{11}_{\mathbf{r}}$, $\{ \tilde{f}^{11}_{\mathbf{r}}, \tilde{f}^{22}_{\mathbf{r}} \}$, and $\{ \tilde{f}^{11}_{\mathbf{r}}, \tilde{f}^{22}_{\mathbf{r}}, \tilde{f}^{12}_{\mathbf{r}} \}$.

\begin{figure}[H]
    \centering
    \includegraphics[width=\textwidth,keepaspectratio]{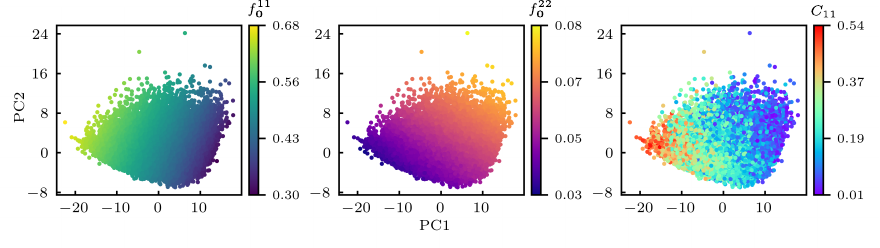}
    \caption{PC representation of all metamaterial unit cells for the combination $\{ \tilde{f}^{11}_{\mathbf{r}}, \tilde{f}^{22}_{\mathbf{r}} \}$ in the space of the first two PCs, colored by different quantities: solid volume fraction $f^{11}_{\mathbf{0}}$ (left), interface volume fraction $\tilde{f}^{22}_{\mathbf{0}}$ (middle), and effective elastic constant $C_{11}$ (right).}
    \label{fig:pca}
\end{figure}

Each set of computed 2-point correlations, derived from a 2D unit cell of $96\times96$ pixels, results in ${96^2=9,216}$ features. This number becomes doubled (18,432) and tripled (27,648) when a combination of two or three sets of correlations is considered, respectively. As discussed in \cref{sec:PCA}, these high-dimensional sets of unwieldy and unmanageable features can be transformed into a low-dimensional space by applying dimensionality reduction. Therefore, we apply PCA to each of the selected combinations to obtain a low-dimensional representation. \cref{fig:pca} illustrates the PC representation of all metamaterial unit cells in the space of the first two PCs, derived from the combination $\{ \tilde{f}^{11}_{\mathbf{r}}, \tilde{f}^{22}_{\mathbf{r}} \}$. The PC representation for this specific combination (i.e., $\{ \tilde{f}^{11}_{\mathbf{r}}, \tilde{f}^{22}_{\mathbf{r}} \}$) is shown because, as will be demonstrated in \cref{sec:ResGPR}, it provides the most valuable statistics for establishing structure–property linkages. In \cref{fig:pca}, the data points are colored by three different quantities, including the solid volume fraction $f^{11}_{\mathbf{0}}$ (left), the interface volume fraction $\tilde{f}^{22}_{\mathbf{0}}$ (middle), and the effective elastic constant $C_{11}$ computed from physics-based simulations (right). Although the adopted feature engineering workflow is unsupervised (i.e., feature selection is independent of the output variable), the data points are colored by the output quantity $C_{11}$ to visualize how the features captured by the PCs relate to the mechanical response. In the left plot of \cref{fig:pca}, PC1 shows a clear correlation with the solid volume fraction. As PC1 decreases from right to left, the solid volume fraction increases. In the middle plot, PC2 appears to exhibit a visible, though weaker, correlation with the interface volume fraction. The right plot shows that $C_{11}$ follows a trend similar to the solid volume fraction along PC1, suggesting that the stiffness of the metamaterial is primarily influenced by the amount of solid material, which is intuitive, as an increased amount of material leads to higher stiffness.

\subsection{Structure-property linkages}
\label{sec:ResGPR}

Once the reduced set of features is obtained, GPR is used to establish the structure–property linkages. We use the FFT-based homogenization scheme introduced in \cref{sec:FFT} to compute the effective elastic constant $C_{11}$. In this setting, the solid regions are modeled as a linear elastic isotropic material with Young’s modulus $E = 1\,\text{GPa}$ and Poisson’s ratio $\nu = 0.3$, while the void regions are assigned $E = 0\,\text{GPa}$ and $\nu = 0$, effectively treating them as empty space with no stiffness contribution. It should be emphasized that the specific value chosen for $E$ does not affect the generality of the results. Since Hooke’s law is linear in $E$, and the second phase in the homogenization problem (i.e., the void) has zero contribution to the stiffness, the effective elastic constant $C_{11}$ is linearly proportional to $E$. Therefore, regardless of the value assigned to $E$, the resulting output represents a normalized elastic constant $C_{11}/E$. This normalization allows the predicted output to be rescaled post prediction to correspond to any desired Young's modulus. For instance, a model trained with $E = 1\,\text{GPa}$ can be readily adapted to predict effective properties for materials with different values of $E$ through a simple scaling operation. In contrast, the effect of Poisson’s ratio $\nu$ is nonlinear. To construct a predictive model applicable to varying values of $\nu$, one must generate data across a range of Poisson’s ratios and include $\nu$ as an additional input variable alongside the PC scores (see, e.g., \cite{danesh2024fft}). In this study, however, we restrict ourselves to a fixed value $\nu = 0.3$, although the approach is  generalizable. After computing the normalized effective elastic constant $C_{11}/E$, we filter out samples with values below $0.01$, as such low stiffness values are not relevant for practical applications. The original dataset from \cite{bastek2023inverse,bastek2023data} included such low-density designs to explore buckling-prone geometries. Applying this filter reduces the dataset size only slightly, from 53,019 to 52,885 samples, eliminating just a small fraction of impractical structures.

Regarding the input parameters for the model, we consider the first 8 PC scores obtained from different combinations of 2-point statistics, including $\tilde{f}^{11}_{\mathbf{r}}$, $\{ \tilde{f}^{11}_{\mathbf{r}}, \tilde{f}^{22}_{\mathbf{r}} \}$, and $\{ \tilde{f}^{11}_{\mathbf{r}}, \tilde{f}^{22}_{\mathbf{r}}, \tilde{f}^{12}_{\mathbf{r}} \}$. Each input feature (i.e., PC score) is standardized to have zero mean and unit variance to improve numerical stability and enhance the performance of GPR model training. We gradually increase the number of PCs from 1 to 8 and train a GPR model for each combination of the mentioned 2-point statistics to identify the model with the highest fidelity. This results in a total of 24 models. The GPR models are implemented using the GPyTorch library, with hyperparameter optimization performed via the Adam optimizer from the scikit-learn library. To ensure the stability of hyperparameter optimization, we perform multiple random initializations and verify that the optimized values converge to consistent solutions. Additionally, a cosine annealing learning rate schedule is employed to promote smooth convergence during training. An $80/20\%$ train/test split is used to evaluate the predictive performance of the trained models for unseen test data. 

\begin{figure}[h]
    \centering
    \includegraphics[keepaspectratio]{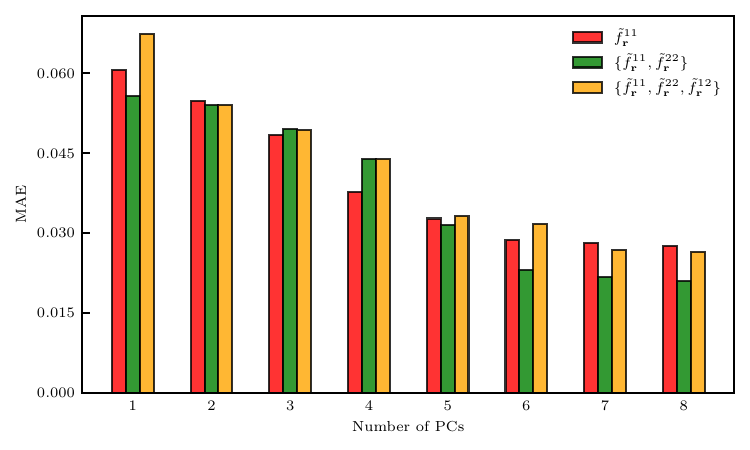}
    \caption{MAE of the test set for GPR models trained with varying numbers of PC scores (from 1 to 8) obtained from three different combinations of 2-point statistics, including $\tilde{f}^{11}_{\mathbf{r}}$, $\{ \tilde{f}^{11}_{\mathbf{r}}, \tilde{f}^{22}_{\mathbf{r}} \}$, and $\{ \tilde{f}^{11}_{\mathbf{r}}, \tilde{f}^{22}_{\mathbf{r}}, \tilde{f}^{12}_{\mathbf{r}} \}$.}
    \label{fig:gpr}
\end{figure}

\cref{fig:gpr} plots the MAE of the test set (computed from \cref{eq:MAE}) for models trained on different numbers of PCs obtained from the various 2-point statistics combinations. The results reveal a consistent decrease in the prediction error as the number of PCs increases, demonstrating the improved predictive power of models incorporating more microstructural features. The inclusion of interface auto-correlation notably improves model performance compared to using only the solid auto-correlation. On the other hand, adding the solid-interface cross-correlation does not appear to further improve the model accuracy. For instance, when using 6 PCs, the MAE decreases to $0.0230$ for the combination including interface auto-correlation (i.e., $\{ \tilde{f}^{11}_{\mathbf{r}}, \tilde{f}^{22}_{\mathbf{r}} \}$), compared to $0.0287$ for the model trained on PCs obtained from the solid auto-correlation alone (i.e., $\tilde{f}^{11}_{\mathbf{r}}$). However, it increases to $0.0317$ when the cross-correlation is also incorporated (i.e., $\{ \tilde{f}^{11}_{\mathbf{r}}, \tilde{f}^{22}_{\mathbf{r}}, \tilde{f}^{12}_{\mathbf{r}} \}$). A saturation in the predictive performance of the models is observed as more PCs are included. For the solid auto-correlation $\tilde{f}^{11}_{\mathbf{r}}$ and the combination $\{ \tilde{f}^{11}_{\mathbf{r}}, \tilde{f}^{22}_{\mathbf{r}} \}$, the models reach a stable level of accuracy with 6 or more PCs. The same trend is observed for the combination $\{ \tilde{f}^{11}_{\mathbf{r}}, \tilde{f}^{22}_{\mathbf{r}}, \tilde{f}^{12}_{\mathbf{r}} \}$, but its performance saturates at 7 PCs or more. These results suggest that the solid and interface auto-correlations (i.e., the combination $\{ \tilde{f}^{11}_{\mathbf{r}}, \tilde{f}^{22}_{\mathbf{r}} \}$) with 6 PC scores form a compact and informative feature set, while the solid–interface cross-correlation does not contribute meaningful additional information for the regression task. It is also worth noting that although the inclusion of interface auto-correlation reduces the MAE by approximately $20\%$ (from $0.0287$ to $0.0230$), the importance of such interface statistics is expected to be even more pronounced in cases involving plasticity or damage, where phase boundaries play a more critical role in mechanical responses such as plastification or damage initiation.

\begin{figure}[h]
    \centering
    \includegraphics[keepaspectratio]{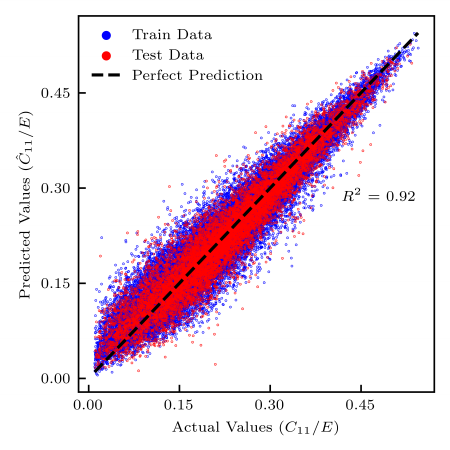}
    \caption{Parity plot showing the predicted versus actual values of the normalized effective elastic constant (i.e., $\hat{C}_{11}/E$ versus $C_{11}/E$) for the model trained with 6 PC scores obtained from the combination $\{ \tilde{f}^{11}_{\mathbf{r}}, \tilde{f}^{22}_{\mathbf{r}} \}$. Results are shown for both the training and test datasets and the diagonal line represents the ideal case of perfect prediction.}
    \label{fig:parity}
\end{figure}

The parity plot, illustrating the predicted versus actual values of the normalized effective elastic constant (i.e., $\hat{C}_{11}/E$ versus $C_{11}/E$) for both the training and test datasets, is shown in \cref{fig:parity} for the optimal model, i.e., the model trained with 6 PC scores obtained from the combination $\{ \tilde{f}^{11}_{\mathbf{r}}, \tilde{f}^{22}_{\mathbf{r}} \}$. The resulting $R^2$ score of approximately $0.92$ is competitive given the lower complexity and parameter efficiency of the GPR model, which achieves strong performance with a small number of kernel hyperparameters compared to the millions of parameters in neural-network-based approaches. While neural networks may achieve higher prediction accuracy (e.g., $R^2$ scores up to $0.99$ as reported in \cite{meyer2022graph,liu2025machine}), such models typically require significantly larger datasets to optimize their extensive parameter space. In contrast, as will be demonstrated in \cref{sec:ResAL}, our GPR approach achieves acceptable saturated accuracy levels with substantially fewer training samples. A detailed comparison with deep learning approaches previously developed for other metamaterials is provided in \cref{sec:ResComparison}.

\subsection{Uncertainty-based active learning}
\label{sec:ResAL}

In this section, we employ the uncertainty-based active learning framework described in \cref{sec:AL} to identify the minimum number of labeled data points required to establish robust structure–property linkages comparable to those trained on the full dataset. As demonstrated in \cref{sec:ResGPR}, the model trained with 6 PCs obtained from the combination $\{ \tilde{f}^{11}_{\mathbf{r}}, \tilde{f}^{22}_{\mathbf{r}} \}$ yields the most accurate predictions. Therefore, we focus solely on this model and disregard other combinations of 2-point correlations and numbers of PCs.

We begin by training a GPR model on 10 randomly selected data points, which is then used to predict the output for the remaining points, referred to as the candidate pool. Subsequently, the point with the highest predictive variance is selected from the candidate pool and added to the training set to retrain an updated GPR model. This iterative procedure is repeated until a specified labeling budget is exhausted or a defined stopping criterion is met. Here, we apply the stopping criterion defined in \cref{eq:stop} to identify the minimum number of observations required for saturated model performance.

\begin{figure}[H]
    \centering
    \includegraphics[keepaspectratio]{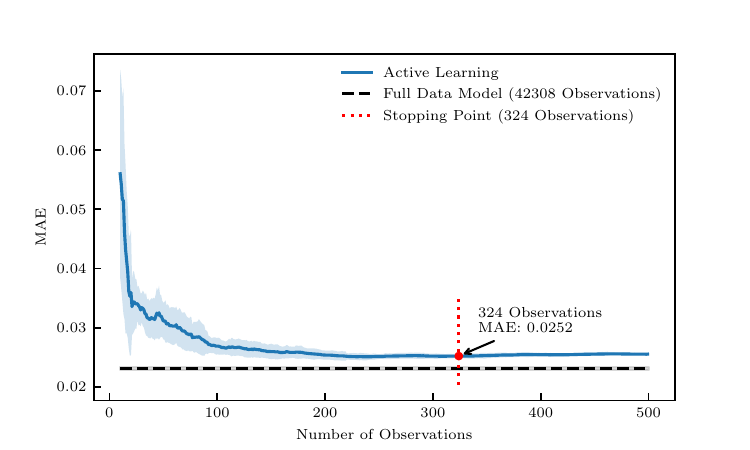}
    \caption{Active learning curve showing MAE over the candidate pool versus the number of labeled observations. The active learning curve corresponds to the mean of 25 independent runs with different random initializations, and the shaded region indicates $\pm 1$ standard deviations. For comparison, the performance of a model trained on the full dataset with an $80/20\%$ train/test split is also shown, averaged over 25 random splits with its own $\pm 1$ standard deviation shaded region.}
    \label{fig:mae_active}
\end{figure}

Since all data points are labeled using the efficient FFT-based homogenization scheme, we are able to evaluate the prediction error using MAE on the entire candidate pool, rather than a hold-out test set, which would typically be the case in real-world applications. \cref{fig:mae_active} plots the MAE over the candidate pool versus the number of labeled observations throughout the active learning process. This curve represents the average of 25 repetitions of the active learning workflow, each with a different random initialization. For comparison, we also include the performance of a GPR model trained on the entire dataset using an $80/20\%$ train/test split, averaged over 25 random splits. Although only $80\%$ of the data is used for training in this case, we refer to this model as the \textit{full data model} for simplicity.

It can be seen that the error decreases sharply during the first 90 iterations, dropping from approximately $0.0560$ for the initial model with $10$ observations to around $0.0269$ for the model with $100$ observations. As active learning continues, the MAE is further reduced to around $0.025$, after which no substantial improvements are observed. By adopting the stopping criterion in \cref{eq:stop} with $Q = 5$ and $\epsilon = 0.0001$, we find that a total of 324 observations (i.e., 10 initial points and 314 active learning iterations) is sufficient for the model to reach saturated performance, with a final MAE of $0.0252$. At this point, the average relative change in MAE over the last 5 iterations falls below $0.0001$. The resulting error is only $0.0021$ higher than that of the model trained on the entire dataset (i.e., $0.0252$ versus $0.0231$), which is acceptable given the significantly smaller training size (324 versus 42,308 points). This number of observations constitutes only about $0.61\%$ of the full dataset, underscoring the effectiveness of the proposed active learning framework.

\begin{figure}[H]
    \centering
    \includegraphics[keepaspectratio]{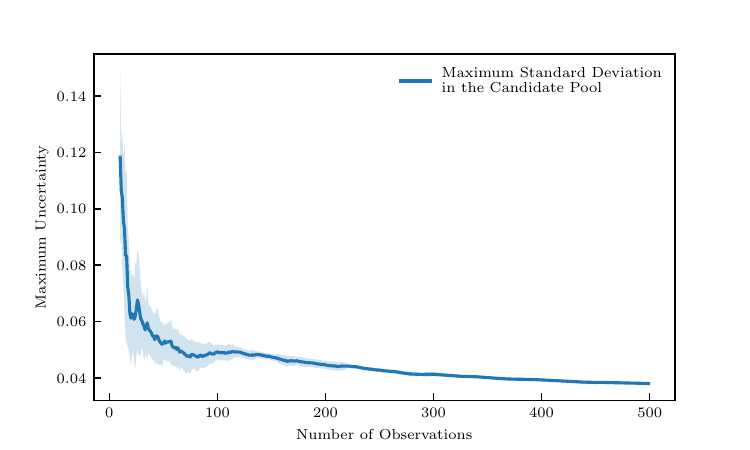}
    \caption{Maximum predictive standard deviation in the candidate pool versus the number of labeled observations throughout the active learning process. The curve corresponds to the mean of 25 independent runs with different random initializations, and the shaded region indicates $\pm 1$ standard deviations.}
    \label{fig:std_active}
\end{figure}

Finally, we can also investigate the prediction uncertainty in the candidate pool by plotting the maximum predictive standard deviation within the pool at each iteration in \cref{fig:std_active}. It is evident that, as the training set is expanded by iteratively adding the most uncertain point, the maximum predictive standard deviation decreases from approximately $0.12$ to about $0.04$. The standard deviation then reaches a plateau near $0.04$, confirming the efficiency of the uncertainty sampling strategy in reducing prediction uncertainty with a relatively small number of observations. Beyond this point, adding more data points does not yield further benefit. This highlights that the adopted active learning workflow enables training a surrogate model with significantly fewer data points, achieving not only predictive accuracy comparable to a model trained on the full dataset but also a substantial reduction in prediction uncertainty.

\subsection{Comparison with deep learning models}
\label{sec:ResComparison}

As a final step of this study, we compare the present workflow with alternative neural-network-based deep learning approaches reported in the literature, in terms of training data efficiency, model complexity, and predictive performance. In particular, we consider the graph neural network (GNN) and the GoogLeNet architecture enhanced with a convolutional block attention module (CBAM), both proposed by \textcite{liu2025machine}, as well as the modified Crystal Graph Convolutional Neural Network (mCGCNN) developed by \textcite{meyer2022graph}. The GNN and GoogLeNet\_CBAM models were trained on 6,531 truss-based lattice structures to predict the elastic modulus. In contrast, the mCGCNN model was trained on 43,505 shell-lattice microstructures to predict a range of effective properties, including thermal, electrical, and magnetic properties, as well as mechanical properties such as the elastic constant $C_{11}$ reported here.

Since the range of the output variables differs between the works considered for comparison and the present study, we normalize the MAE (i.e., \cref{eq:MAE}) by the range of the output variable in each dataset (i.e., the difference between the maximum and minimum output values), and refer to the resulting quantity as the normalized mean absolute error (nMAE). For a consistent comparison, \cref{tab:comparison} summarizes the dataset sizes, the fraction of dataset size used for training (training set fraction), model complexities in terms of number of parameters, and the nMAE achieved by the different models. We also include our models trained on the full dataset and achieved through active learning. Notably, the dataset used in the present study comprises 52,885 samples, making it the largest among the compared studies. Although the nMAE of our active learning model ($4.7\%$) is slightly higher than the errors achieved by the neural-network-based models, it is important to highlight that our model attains this level of accuracy by using only $0.61\%$ of the entire dataset for training, whereas the other approaches used $90\%$ of their datasets. Furthermore, our GPR model is remarkably compact, involving only eight model parameters: an output variance factor, a noise factor, and 6 characteristic length scales corresponding to 6 PC scores, i.e., $\boldsymbol{\theta} = \left( \sigma_f, \sigma_n, \ell_1, \ell_2, \dots, \ell_6 \right)$. In contrast, the other deep learning approaches require tens or hundreds of thousands of parameters to achieve such high levels of predictive accuracy. These results underscore the efficiency, simplicity, and strong predictive capabilities of the proposed framework, particularly when combined with active learning for data selection.

\begin{table}[h]
\caption{Comparison with different available deep learning models.}
\label{tab:comparison}
\centering
\begin{tabularx}{\textwidth}{l c c c c c}
\hline
Model & \makecell{Dataset\\Size} & \makecell{Train Set\\Fraction} & \# Parameters & nMAE & \makecell{Predicted\\Property} \\
\hline
GNN \cite{liu2025machine} & 6,531 & 0.9 & 44,401 & $2.2\%$ & Elastic modulus \\
GoogLeNet\_CBAM \cite{liu2025machine} & 6,531 & 0.9 & 1,038,755 & $1.3\%$ & Elastic modulus \\
mCGCNN \cite{meyer2022graph} & 43,505 & 0.9 & $\sim\,$230,000 & $2.2\%$ & $C_{11}$ \\
Present (Full data model) & 52,885 & 0.8 & 8 & $4.3\%$ & $C_{11}$ \\
Present (Active learning model) & 52,885 & 0.0061 & 8 & $4.7\%$ & $C_{11}$ \\
\hline
\end{tabularx}
\end{table}

\section{Conclusions}

In this work, we employed the MKS framework to establish structure–property linkages for a large dataset of periodic 2D stochastic metamaterial unit cells. An FFT-based homogenization scheme was used to efficiently compute the effective elastic constant $C_{11}$ (i.e., the output of interest) for the entire dataset. A low-dimensional representation of the microstructures was obtained by first computing 2-point spatial correlations and subsequently applying PCA. To investigate the role of the solid–void interface, we extracted the interface for each unit cell via a 2D convolution operation and included its 2-point statistics in combination with the correlations of the solid phase to assess its effect on model fidelity. Different combinations of solid and interface correlations were then considered to train GPR models. Each model was evaluated across varying numbers of PC scores to identify the most accurate surrogate with the fewest input parameters. It was found that the combination of solid and interface auto-correlations with 6 PC scores yielded the most accurate model. However, the inclusion of the solid–interface cross-correlation did not lead to noticeable improvements in model accuracy, suggesting it provides limited additional information for the regression task. Subsequently, an uncertainty-based active learning workflow was employed to determine the minimum number of labeled data points required to train a high-fidelity surrogate. We demonstrated that using only approximately $0.61\%$ of the entire dataset, it is possible to achieve predictive performance comparable to that of a model trained on the full dataset. This highlights the data efficiency of the proposed approach, which achieves competitive accuracy compared to more data-intensive neural-network-based methods. Moreover, the use of GPR naturally provides prediction uncertainty, which is not inherently available in neural-network-based models and typically requires additional treatment.

For future studies, the proposed workflow can be extended to more complex material behaviors, such as plasticity and damage \cite{brepols2017gradient,brepols2020gradient}. As demonstrated in previous works \cite{montes2018reduced,venkatraman2020reduced}, the influence of interface statistics on the fidelity of surrogate models is expected to be even more pronounced in these regimes, further motivating continued investigation. Given the importance of multiscale modeling in metamaterials \cite{kochmann2019multiscale,danesh2023challenges,danesh2025two}, a potential direction for future work is to develop a two-scale approach in which finite element analysis is employed for the macroscale problem and the MKS framework is used to model microscale unit cells, as previously adopted for composite materials \cite{al2012multi}. Moreover, the MKS framework has already been extended to inverse stochastic microstructure design and applied to woven composites \cite{generale2024inverse}, motivating the development of an inverse design approach to identify metamaterial unit cell configurations that yield target mechanical responses. Eventually, although localization has been explored for materials with varying stiffness contrast using neural networks \cite{yang2019establishing,harandi2024spectral,kelly2025thermodynamically} or the MKS localization framework \cite{fast2011formulation,yabansu2014calibrated,de2019localization}, challenges in high contrast scenarios necessitates studying localization for metamaterial unit cells with infinite stiffness contrast. Such an investigation would seek to accurately predict not only homogenized properties but also full-field solutions, thereby enhancing the predictive capabilities of the present methodology.

\newpage
\section*{Acknowledgments}

\noindent
\includegraphics[width=0.15\linewidth]{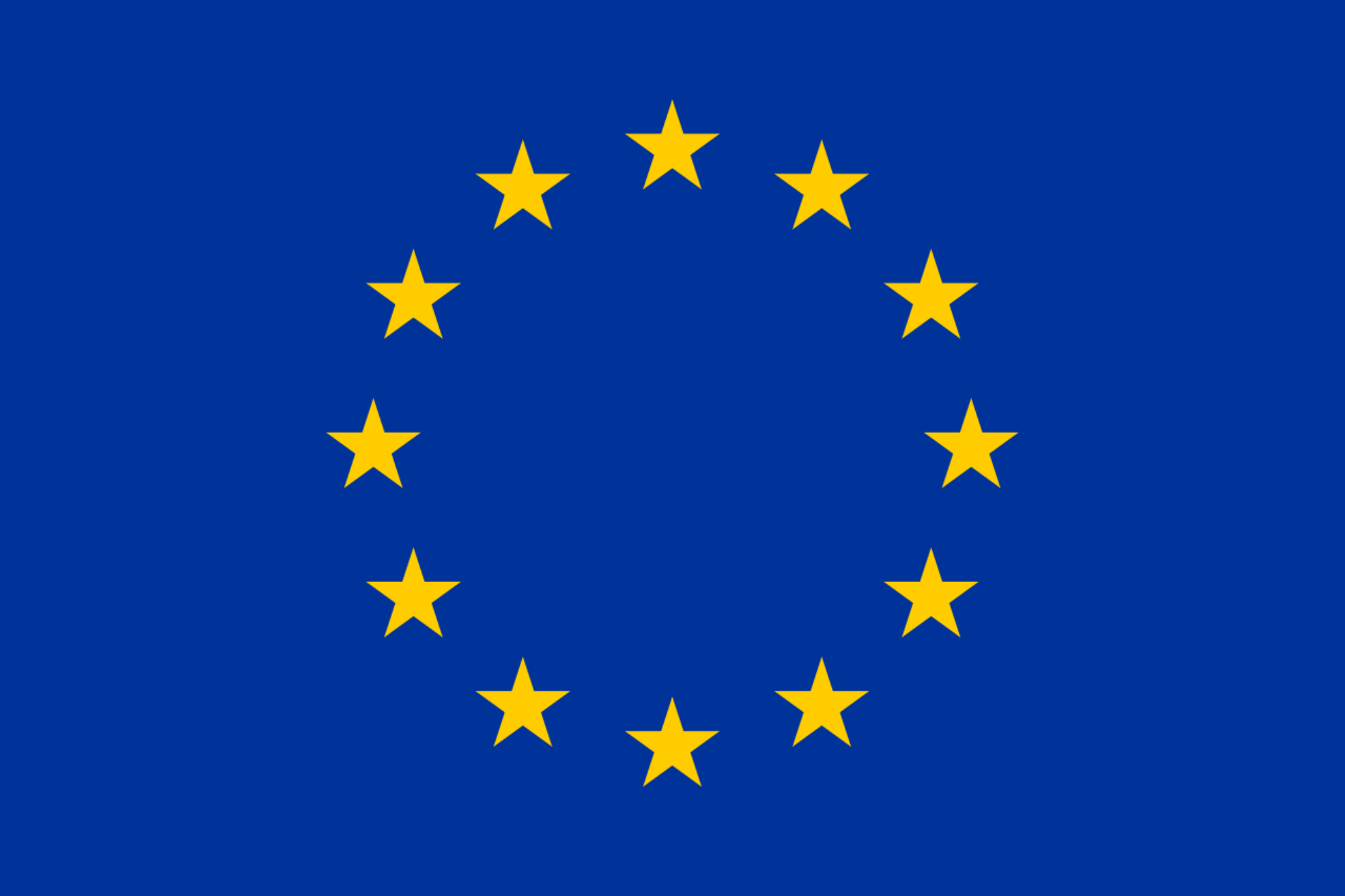}

\noindent
H. Danesh, T. Brepols, and S. Reese: The authors gratefully acknowledge the funding from the European Union’s Horizon 2020 research and innovation program under the Marie Skłodowska-Curie grant agreement No 956401 (XS-Meta).
M. Annamaraju and S. Kalidindi: Research was sponsored by the Army Research Laboratory and was accomplished under Cooperative Agreement Number W911NF-22-2-0106. The views and conclusions contained in this document are those of the authors and should not be interpreted as representing the official policies, either expressed or implied, of the Army Research Laboratory or the U.S. Government. The U.S. Government is authorized to reproduce and distribute reprints for Government purposes notwithstanding any copyright notation herein.

% \bibliographystyle{unsrt}
% \bibliographystyle{apalike}
% \bibliography{Bib}

\section*{Data Availability}
The structure–property dataset generated in this work, containing binary topologies of metamaterial unit cells and their corresponding normalized effective elastic constants, is publicly available at \cite{danesh2025data}:  
\url{https://doi.org/10.5281/zenodo.15302946}. The unit cell topologies were extracted from the original dataset provided in \cite{bastek2023data}, which is licensed under the \href{https://creativecommons.org/licenses/by/4.0/}{Creative Commons Attribution 4.0 International License}.

\printbibliography

\end{document}